\documentclass[submission,publicdomain,creativecommons]{eptcs}
\providecommand{\event}{PxTP 2017} % Name of the event you are submitting to
\usepackage{breakurl}             % Not needed if you use pdflatex only.
\usepackage{underscore}           % Only needed if you use pdflatex.

\usepackage[utf8x]{inputenc}

\usepackage{amsmath}
\usepackage{amssymb}
\usepackage{upgreek}
\usepackage{color}
\definecolor{keywordcolor}{rgb}{0.7, 0.1, 0.1}   % red
\definecolor{tacticcolor}{rgb}{0.1, 0.2, 0.6}    % blue
\definecolor{commentcolor}{rgb}{0.4, 0.4, 0.4}   % grey
\definecolor{symbolcolor}{rgb}{0.0, 0.1, 0.6}    % blue
\definecolor{sortcolor}{rgb}{0.1, 0.5, 0.1}      % green

\usepackage{listings}
\usepackage{scalefnt}
\usepackage{relsize}
\usepackage{hyperref}

\hypersetup{
    colorlinks=true,
    citecolor=blue,
    linkcolor=blue,
    urlcolor=blue
}
\usepackage{verbatim}% http://ctan.org/pkg/verbatim

\newlength\myverbindent 
\makeatletter
\newcommand{\verbatimfont}[1]{\def\verbatim@font{#1}}%
\def\verbatim@processline{%
  \hspace{\myverbindent}\the\verbatim@line\par}
\makeatother
\verbatimfont{\fontfamily{qcr}\selectfont\scalefont{.93}}

\newcommand{\mm}[1]{\mbox{\fontfamily{qcr}\selectfont\scalefont{.93}#1}}%{\texttt{#1}}
\newcommand{\lean}[1]{\lstinline[language=lean]{#1}}
\usepackage{url}

\title{An Extensible Ad Hoc Interface between \\ Lean and Mathematica}
\author{Robert Y. Lewis
\institute{Carnegie Mellon University \\ Pittsburgh, PA, USA}
\email{rlewis1@andrew.cmu.edu}
}

\def\titlerunning{An Extensible Ad Hoc Interface between Lean and Mathematica}
\def\authorrunning{R.Y. Lewis}
\begin{document}
\maketitle

\begin{abstract}
We implement a user-extensible ad hoc connection between the Lean proof assistant and the computer algebra system Mathematica. By reflecting the syntax of each system in the other and providing a flexible interface for extending translation, our connection allows for the exchange of arbitrary information between the two systems. We show how to make use of the Lean metaprogramming framework to verify certain Mathematica computations, so that the rigor of the proof assistant is not compromised.
\end{abstract}

\section{Introduction}
\label{section:introduction}

Many researchers have noted the disconnect between computer algebra and interactive theorem proving. In the former, one typically values speed and flexibility over absolute correctness. To be more efficient or user-friendly, a computer algebra system (CAS) may blur the distinction between polynomial objects and polynomial functions, assume that sufficiently small terms at the end of a series are zero, or resort to numerical approximations without warning. Such simplifying assumptions can make sense in the context of computer algebra; the capabilities of these systems make them indispensable tools to many working mathematicians. These assumptions, though,
are antithetical to the goals of interactive theorem proving (ITP), where every inference must be justified by appeal to some logical principle. The strict logical requirements and lack of many familiar algorithms discourage many mathematicians from using proof assistants.

Integrating computer algebra into proof assistants is one way to reduce this barrier to entry, and bridges between the two types of systems have been built in a variety of ways. We contribute another such bridge,  between the proof assistant Lean \cite{demoura:14} and the computer algebra system Mathematica \cite{Wolfram2015}. Our connection is inspired by the architecture described by Harrison and Th\'ery \cite{Harrison1998}. By integrating with the Lean metaprogramming framework, our design allows users to verify results from the CAS without leaving the Lean environment. Since Mathematica is one of the most commonly used computer algebra systems, and a user with knowledge of the CAS can extend the capabilities of our link, we hope that the familiarity will lead to wider use. 

Our link separates the steps of communication, semantic interpretation, and verification: there is no a priori restriction on the type of information that can be shared between the systems. With the proof assistant in the ``master'' role, Lean expressions are exported to Mathematica, where they can be interpreted and manipulated. The results are then imported back into Lean and reinterpreted. Using Lean's metaprogramming framework, one can write scripts that verify properties of the translated results. This style of interaction, where verification happens on a per-case basis after the computation has ended, is called \emph{ad hoc}. 

By performing calculations in Mathematica and verifying the results in Lean, we relax neither the rigor of the proof assistant nor the efficiency of the CAS. The CAS can alternatively be used as an untrusted oracle, or can play a purely informative role, where its output does not appear in the final proof term. This range of possibilities is intended to make our link attractive to multiple audiences. The working mathematician, who balks at the restrictions imposed by a proof assistant, may find that full access to a familiar CAS is worth the tradeoff in trust. Industrial users are often happy to trust both large-kernel proof assistants and computer algebra systems; the rigor of Lean with Mathematica as an oracle falls somewhere in between. And certifiable algorithms are still available to users who demand complete trust.

The source for this project, and supplementary documentation, is available at \url{http://www.andrew.cmu.edu/user/rlewis1/leanmm/}. In this paper, we use \lean{Computer Modern} for Lean code and \mm{TeX} \mm{Gyre} \mm{Cursor} for Mathematica code. We begin by describing some of the salient features of the two systems. Section \ref{section:translation} discusses the translation of Lean expressions into semantically similar Mathematica expressions, and vice versa. Section \ref{section:interface} describes further details of the implementation. In {Section \ref{section:verification}} we give examples of the link in action. We conclude with a discussion of related and future work.

\section{System descriptions: Lean and Mathematica}
\label{section:systems}

\subsection{Lean}
\label{subsection:systems:lean}
Lean is a proof assistant being developed at Microsoft Research \cite{demoura:14}. 
Written in C++, the system is highly performant. Lean has been designed from the beginning to support strong automation; it aims to eventually straddle the line between an interactive theorem prover with powerful automation, and an automated theorem prover with a verified code base and interactive mode. 

Lean is based on the Calculus of Inductive Constructions (CIC), an extension of the lambda-calculus with dependent types and inductive definitions. There is a non-cumulative hierarchy of type universes \lean{Sort u}, \lean{u ≥ 0}, with the abbreviations \lean{Prop = Sort 0} and \lean{Type u = Sort (u+1)}.
The bottom level \lean{Prop} is impredicative and proof-irrelevant. We refer readers to \cite{Coquand1988}, \cite{Coquand1990}, and \cite{demoura:14} for more details about the CIC and Lean's implementation. 

Lean's standard library uses type classes to implement an abstract algebraic hierarchy. Arithmetic operations, such as $+$ and $*$, and numerals are generic over types that instantiate the appropriate classes. As an example, the addition operator has the signature
%\vspace{-.3em}
\begin{lstlisting}[language=lean]
add {u} : Π {A : Type u} [has_add A], A → A → A.
\end{lstlisting}
%\vspace{-.3em}
The notation \lean{\{A : Type u\}} denotes that the argument \lean{A} is an implicit variable, meant to be inferred from further arguments; \lean{has_add : Type u → Type u} is a type class, and the notation \lean{[has_add A]} denotes that a term of that type is to be inferred using type class resolution. The universe argument \lean{u} indicates that \lean{add} is parametric over one universe level.

The dependently typed language implemented in Lean is flexible enough to serve as its own \emph{metaprogramming language} \cite{lean:popl17}. Data types and procedures implemented in Lean's underlying C++ code base are exposed as constants, using the keyword \lean{meta} to mark a distinction between the object language and this extension. Expressions can be evaluated in the Lean virtual machine, which replaces these constants with their underlying implementation. Meta-definitions permit unbounded recursion but are otherwise quite similar to standard definitions.

Combined with the declaration of the types \lean{pexpr} and \lean{expr}, which expose the syntax of Lean (pre-)expressions in Lean itself, and \lean{tactic_state}, which exposes the environment and goals of a tactic proof, this metaprogramming framework allows users to write complex procedures for constructing proofs. A term of type \lean{tactic A} is a function \lean{tactic_state → tactic_result A}, where a result is either a success (pairing a new \lean{tactic_state} with a term of type \lean{A}) or a failure. Proof obligations can be discharged by terms of type \lean{tactic unit}; such a term is executed in the Lean virtual machine to transform the original \lean{tactic_state} into one in which all goals have been instantiated. More generally, we can think of a term of type \lean{tactic A} as a program that attempts to construct a term of type \lean{A}, while optionally changing the tactic state. 

When writing tactics, the command \lean{do} enables Haskell-like monadic syntax. For example, the following tactic returns the number of goals in the current tactic state. 
The type of \lean{get_goals} is \lean{tactic (list expr)}, where \lean{list} is the standard (object-level) type defined in the Lean library.
%%\vspace{-.3em}
\begin{lstlisting}[language=lean]
meta def num_goals : tactic nat :=
do gs ← get_goals,
   return (length gs)
\end{lstlisting}

Lean allows the user to tag declarations with \emph{attributes}, and provides an interface \lean{name → tactic (list name)}
to retrieve a list of declarations tagged with a certain attribute. 

Many features and subtleties of the metaprogramming framework are discussed in \cite{lean:popl17}. In closing, we note how the framework is used for this project. 

We define the function \lean{mm_form_of_expr : expr → string} recursively on the type \lean{expr} to represent Lean syntax in Mathematica. We also define a function \lean{mathematica.execute : string → tactic mmexpr}. This function, which uses Lean's IO monad to communicate with external programs, passes the input string to Mathematica and returns a Lean expression encoding Mathematica's output. The type \lean{mmexpr}, which represents Mathematica's term structure, is described in Section \ref{subsection:translation:ml}. The program \lean{expr_of_mmexpr : mmexpr → tactic expr} and variants search the context for attributed translation rules, and try to apply these rules to convert the Mathematica expression into a meaningful Lean expression. Finally, various tactics are defined to make use of these results.

\subsection{Mathematica}
\label{subsection:systems:mathematica}
Mathematica is a popular symbolic computation system developed at Wolfram Research, implementing the Wolfram Language \cite{Wolfram2015}. Along with support for a vast range of mathematical computations, Mathematica includes collections of data of various types and tools for manipulating this data.

Mathematica provides comprehensive tools for rewriting and solving polynomial, trigonometric, and other classes of equations and inequalities; solving differential equations, both symbolically and numerically; computing derivatives and integrals of various types; manipulating matrices; performing statistical calculations, including fitting and hypothesis testing; and reasoning with classes of special functions. 

This large library of functions is one reason to choose Mathematica for our linked CAS. Another reason is its ubiquity: Mathematica is frequently used in undergraduate mathematics and engineering curricula. Lean beginners who are accustomed to Mathematica do not need to learn a new CAS language for the advanced features of this link. 

For those unfamiliar with the syntax of the Wolfram Language, we note some features and terminology that will help to understand the code fragments in this paper.
\begin{itemize}
 \item Function application is written using square brackets, e.g. \mm{Plus[x, y]}. Many functions are variadic: that is, we can also write \mm{Plus[x, y, z]}. Of course, we can use common notation like \mm{x + y + z} instead.
 \item Alternatively, we can write unary function application in postfix form: 
 
 \mm{x\^{}2 - 2x + 1 // Factor} is equivalent to \mm{Factor[x\^{}2 - 2x + 1]}. 
 \item In the expression \mm{Plus[x, y]}, we refer to \mm{Plus} as the \emph{head symbol} and \mm{x} and \mm{y} as the \emph{arguments}. In general, non-numeric atoms like \mm{x} and \mm{y} are called \emph{symbols}.
 \item There is no strong distinction between defined and undefined symbols. The user is free to introduce a new symbol and use it at will. The computational behavior of a head symbol can be fully or partially defined via pattern matching rules, such as \mm{F[x_,y_] := x + y}; the underscores indicate that \mm{x_} and \mm{y_} are patterns.
 \item The Wolfram Language is untyped, so head symbols such as \mm{Plus} and \mm{Factor} can be applied to any argument or sequence of arguments. Evaluation is often restricted to certain patterns: \mm{Plus[2, 3]} will evaluate to \mm{5}, but \mm{Plus[Factor, Plus]} will not reduce. Nevertheless, both are well-formed Mathematica expressions.
\end{itemize}
\vspace{-1em}

\section{The translation procedure}
\label{section:translation}
Our bridge is used to import information from Mathematica into Lean, usually about some particular Lean expression. The logical foundations and semantics of the two systems are quite different, and we should not expect a perfect correspondence between the two. However, in many situations, an expression in Lean has a counterpart in Mathematica with a very similar intended meaning. We can exploit these similarities by ignoring the unsoundness of the translations in both directions and attempting to verify, post hoc, that the resulting expression has the intended properties.

As a running toy example, suppose we want to show in Lean that 
%\vspace{-.3em}
\begin{lstlisting}[language=lean]
x : real ⊢ x^2 - 2x + 1 ≥ 0. 
\end{lstlisting}
%\vspace{-.3em}
Factoring the left-hand side of the inequality makes this a one-step proof (assuming we've proved that squares are nonnegative). It is nontrivial to write a reliable and efficient polynomial factoring algorithm, but luckily, one is implemented in Mathematica. So we would like to do the following:
\begin{enumerate}
\item Transform the Lean representation of $x^2 - 2x + 1$ into Mathematica syntax.
\item Interpret this into the Mathematica representation of the same polynomial.
\item Use Mathematica's \mm{Factor} function to factor the polynomial.
\item Transform this back into Lean syntax, and interpret it as a Lean polynomial.
\item Verify that the new expression is equal to the old.
\item Substitute this equality into the goal.
\end{enumerate}

We discuss steps 5 and 6 in section \ref{section:verification}; since checking that a polynomial has been factored correctly is much easier than factoring it in the first place, these are handled easily by simplification and rewriting.
%Step 6 is easily achieved with Lean's \lean{rewrite} tactic. Lean's simplifier handles step 5 -- checking that a polynomial has been factored correctly is much easier than factoring it in the first place. 
And, once we have a valid Mathematica expression, step 3 is trivial. In this section we describe steps 1, 2, and 4.

It is worth emphasizing the modularity and extensibility of this approach. Both directions of translation are handled independently, and the translation rules can be extended or changed at will. Translation rules may be arbitrarily complex. Users may choose to use alternate verification procedures, or to forego the verification step entirely.
%\vspace{-.3em}

\subsection{Translating Lean to Mathematica}
\label{subsection:translation:lm}

The Lean expression grammar is presented (in Lean syntax) in Figure~\ref{figure:leankinds}; we elaborate below. For the sake of brevity, we will not discuss the implementations of \lean{name}, \lean{level}, or \lean{binder_info}. The type is marked with the keyword \lean{meta} because, during evaluation, the Lean virtual machine replaces terms of type \lean{expr} with the kernel's expression datatype.
\begin{figure}[t!]
\begin{lstlisting}[language=lean]
meta inductive expr
| var         : nat → expr
| sort        : level → expr
| const       : name → list level → expr
| mvar        : name → expr → expr
| local_const : name → name → binder_info → expr → expr
| app         : expr → expr → expr
| lam         : name → binder_info → expr → expr → expr
| pi          : name → binder_info → expr → expr → expr
| elet        : name → expr → expr → expr → expr
| macro       : macro_def → list expr → expr
\end{lstlisting}
\caption{Lean expression kinds}
\label{figure:leankinds}
\end{figure}

Each Lean expression exists in an environment, which contains the names, types, and definitions of previous declarations. The \lean{const} kind accesses a previous declaration, instantiated to particular universe levels if the declaration is parametric. In addition to declarations in its environment, an expression may refer to its local context, which contains variables and hypotheses of kind \lean{local_const}. In the toy example introduced above, \lean{x} is a local constant. A local constant has a unique name, a formatting name, and a type. 

The expression kinds \lean{lam} and \lean{pi} respectively represent lambda-abstraction and the dependent function type. (Non-dependent function types are degenerate cases of pi types.) Each contains a name for the bound variable, the type of the variable, and the expression body. Bound variables of kind \lean{var} are anonymous within the body, being represented by De Bruijn indices \cite{McBride2004}. Application of one expression to another is represented by the \lean{app} kind.

Type universes are implemented by the expression kind \lean{sort}. Metavariables represent placeholders in partially constructed expressions; the \lean{mvar} kind holds the name and type of the placeholder. Let expressions (\lean{elet}) bind a named variable with a type and value within a body. We do not describe macro expressions, as they are not supported by our link.

To represent this syntax in Mathematica, we define \lean{mathematica_form_of_expr : expr → string} by recursion over the \lean{expr} datatype. We associate a Mathematica head symbol \mm{LeanVar}, \mm{LeanSort}, \mm{LeanConst}, etc.\ to each constructor of \lean{expr}. Names, levels, lists of levels, and binder information are also represented.

Some of the information contained in a Lean expression has little plausible use in Mathematica, or is needlessly verbose: for example, it is hard to contrive a scenario in which the full structure of a Lean \lean{name} is used in the CAS. Nonetheless, we do not strip any information at this stage, to preserve the property that an expression reflected into and immediately back from Mathematica should translate to the original expression without any additional information. 
 
In our running example, we work on the expression $x^2 - 2x + 1$. The fully-elaborated Lean expression and its Mathematica representation are too long to print here, but they can be viewed in the supplementary documentation;
we consider the more concise example of $x+x$. 
If we use strings to stand in for terms of type \lean{name}, natural numbers in place of universe levels, and the string \lean{"bi"} in place of the default \lean{binder_info} argument, and we abbreviate
%\vspace{-.3em}
\begin{lstlisting}[language=lean]
𝒳 := local_const "17.27" "x" "bi" (const "real" []),
\end{lstlisting}
%\vspace{-.3em}
the full form of $x+x$ is
%\vspace{-.3em}
\begin{lstlisting}[language=lean]
app (app (app (app (const "add" [0]) (const "real" [])) 
              (const "real.has_add" [])) 𝒳) 𝒳. 
\end{lstlisting}
%\vspace{-.3em}
The corresponding Mathematica expressions are
%\vspace{-.3em}
\begin{verbatim}
X := LeanLocal["17.27", "x", "bi", LeanConst["real", {}]]

LeanApp[LeanApp[LeanApp[LeanApp[LeanConst["add", {0}], 
  LeanConst["real", {}]], LeanConst["real.has_add", {}]], 
  X], X].
\end{verbatim}
%%\vspace{-.3em}

(In these expressions, ``17.27'' is a unique name for the variable \lean{x}, used only internally.)

Since the head symbols \mm{LeanApp}, \mm{LeanConst}, etc.\ are uninterpreted in Mathematica, this representation is not yet useful. We wish to exploit the fact that many Lean terms have semantically similar counterparts in Mathematica. For instance, the Lean constants \lean{add} and \lean{mul} behave similarly to the Mathematica head symbols \mm{Plus} and \mm{Times}; both systems have notions of application, although they handle the arity of applications differently; and Mathematica's concept of a ``pure function" is analogous to lambda-abstraction in Lean.

We thus define a translation function \mm{LeanForm} in Mathematica that attempts to interpret the syntactic representation. Mathematica functions are typically defined using pattern matching. The \mm{LeanForm} function, then, will look for familiar patterns (e.g. \lean{add A h x y}, in Mathematica syntax) and rewrite them in translated form (e.g. \mm{Plus[LeanForm[x], LeanForm[y]]}). Users can easily extend this translation function by asserting additional equations; a default collection of equations is loaded automatically.

For our factorization example, we want to convert Lean arithmetic to Mathematica arithmetic. Among other similar rules, we will need the following:
%\vspace{-.3em}
\begin{verbatim}
LeanForm[LeanApp[LeanApp[LeanApp[LeanApp[LeanConst["add",_],
_], _], x_], y_]] := Inactive[Plus][LeanForm[x],LeanForm[y]]
\end{verbatim}   

Note that this pattern ignores the type argument and type-class instance in the Lean term. These arguments are irrelevant to Mathematica and can be inferred again by Lean in the back-translation.
We block Mathematica's computation with the \mm{Inactive} head symbol; otherwise, Mathematica would eagerly simplify the translated expression, which can be undesirable. The function \mm{Activate} strips these annotations and allows reduction.

Numerals in Lean are type-parametric and are represented using the constants \lean{zero}, \lean{one}, \lean{bit0}, and \lean{bit1}. To illustrate, the type signature of the latter is
%\vspace{-.3em}
\begin{lstlisting}[language=lean]
bit1 {u} : Π {A : Type u}, [has_add A] → [has_one A] → A → A
\end{lstlisting}
%%\vspace{-.3em}
and the numeral 6 is represented as \lean{bit0 (bit1 one)}; the type of this numeral is expected to be inferable from context. We can use rules similar to the above to transform Lean numerals into Mathematica integers:
%\vspace{-.3em}
\begin{verbatim}
LeanForm[LeanApp[LeanApp[LeanApp[LeanApp[
  LeanConst["bit1", _], _], _], _], t_]] := 2*LeanForm[t]+1.
\end{verbatim}
%\vspace{-.3em}

Applying \mm{LeanForm} will not necessarily remove all occurrences of the head symbols \mm{LeanApp}, \mm{LeanConst}, etc. This is not a problem: we only need to translate the ``concepts'' with equivalents in Mathematica. Unconverted subterms -- for instance \mm{X}, which contains applications of \mm{LeanLocal} and \mm{LeanConst} -- will be treated as uninterpreted constants by Mathematica, and the back-translation described below will return them to their original Lean form.

In our running example (keeping the abbreviation \mm{X}), applying the \mm{LeanForm} and \mm{Activate} functions produces the expression
%\vspace{-.3em}
\begin{verbatim}
Plus[1,Times[-2, X], Power[X, 2]].
\end{verbatim}
%\vspace{-.3em}
Applying \mm{Factor} produces \mm{Power[Plus[-1, X], 2]}. 

The expression
%\vspace{-.3em}
\begin{verbatim}
X := LeanLocal["17.27", "x", "bi", LeanConst["real",{}]]
\end{verbatim}
%\vspace{-.3em}
has been treated as a constant throughout the process, and contains information that will rarely if ever be of use in Mathematica. The excess information does little harm here, 
%(besides making this paper difficult to format), 
but in more complex situations, carrying around this excess information can be unwieldy. We provide Mathematica functions \mm{LeanCollapse} and \mm{LeanInflate} to reduce this excess baggage during computation.

\subsection{Translating Mathematica to Lean}
\label{subsection:translation:ml}

Mathematica expressions are composed of various atomic number types, strings, symbols, and applications, where one expression is applied to a list of expressions. We represent this structure in Lean with the data type \lean{mmexpr} (Figure~\ref{figure:mmexpr}).
\begin{figure}[!t]
\begin{lstlisting}[language=lean]
inductive mmexpr 
| sym   : string → mmexpr
| mstr  : string → mmexpr
| mint  : int → mmexpr 
| app   : mmexpr → list mmexpr → mmexpr
| mreal : float → mmexpr 
\end{lstlisting}
\caption{Mathematica expression kinds}
\label{figure:mmexpr}
\end{figure}

The result of a Mathematica computation is reflected into Lean as a term of type \lean{mmexpr}. This is analogous to the original export of our Lean expression into Mathematica; it remains to interpret it as something meaningful.

A \emph{pre-expression} in Lean is a term where universe, implicit, and inferable arguments are omitted.
It is not expected to type check, but one can try to convert it into a type-correct term via elaboration. For instance, the pre-expression \lean{```(add nat.one nat.one)} elaborates to \lean{add.\{0\} nat nat.has_add nat.one nat.one}. The notation \lean{```(...)} instructs Lean's parser to interpret the quoted text as a term of type \lean{pexpr}. Pre-expressions share the same structure as expressions; in fact, the types \lean{expr} and \lean{pexpr} are isomorphic.

Most Mathematica expressions correspond to pre-expressions: they may be type-ambiguous, and contain less information than their Lean counterparts. 
Thus we normally expect to interpret terms of type \lean{mmexpr} as pre-expressions, and to use the Lean elaborator to turn them into full expressions. However, in rare cases an \lean{mmexpr} may already correspond to a full expression: the unmodified representation of a Lean expression, sent back into Lean, should interpret as the original expression. We provide two extensible translation functions, \lean{expr_of_mmexpr} and \lean{pexpr_of_mmexpr}, to handle both of these cases. Since the implementations are similar, we focus on the latter here.

The function \lean{pexpr_of_mmexpr : trans_env → mmexpr → tactic pexpr} takes a translation environment and an \lean{mmexpr}, and, using the attribute manager, attempts to return a pre-expression. (Since the tactic monad includes failure, the process may also fail if no interpretation is found.) Interpreting strings as pre-expressions, or, indeed, as expressions, is straightforward. Since Mathematica
ints may be used to represent numerals in many different Lean types,
expressions built with \lean{mint} are interpreted as untyped numeral pre-expressions. 

The \lean{sym} and \lean{app} cases are more complex: this part of the translation procedure is extensible by the user. We define three classes of translation rules:
\begin{itemize}
\item A sym-to-pexpr rule, of type \lean{string × pexpr}, identifies a particular Mathematica symbol with a particular pre-expression. For example, the rule \lean{("Real", ```(real))} instructs the translation to replace the Mathematica symbol \mm{Real} with the Lean pre-expression \lean{const "real"}.

\item A keyed app-to-pexpr rule is of type \lean{string × (trans_env → list mmexpr → tactic pexpr)}. When the procedure encounters an \lean{mmexpr} of the form \lean{app (sym head) args} -- that is, the Mathematica head symbol \mm{head} applied to a list of arguments \mm{args} -- it will try to apply all rules that are keyed to the string \lean{head}. The rules for interpreting arithmetic expressions follow this pattern: a rule keyed to the string \lean{"Plus"} will interpret \mm{Plus[t\textsubscript{1}, ..., t\textsubscript{\emph{n}}]} by folding applications of \lean{add} over the translations of \mm{t\textsubscript{1}} through \mm{t\textsubscript{\emph{n}}}.

\item An unkeyed app-to-pexpr rule is of type \lean{trans_env → mmexpr → list mmexpr → tactic pexpr}. If the head of the application is a compound expression, or if no keyed rules execute successfully, the translation procedure will try unkeyed rules. One such rule attempts to translate the head symbol and arguments independently, and fold application over these translations.
\end{itemize}

Rules of these three types can be declared by the user and tagged with the corresponding attribute. The translation procedure uses Lean's caching attribute manager to collect relevant rules at runtime.

Returning to our example, we have translated the expression \lean{x^2 - 2x + 1} and factored the result, to produce 
\mm{Power[Plus[-1, X], 2]}. This is reflected as the Lean \lean{mmexpr}
%\vspace{-.3em}
\begin{lstlisting}[language=lean]
app (sym "Power") [app (sym "Plus") [mint -1, X], mint 2],
\end{lstlisting}
%\vspace{-.3em}
where
%\vspace{-.3em}
\begin{lstlisting}[language=lean]
X := app (sym "LeanLocal") [str "17.27", str "x", str "bi", 
                        app (sym "LeanConst") [str "real", []]].
\end{lstlisting}
%\vspace{-.3em}

Applying \lean{pexpr_of_mmexpr} produces the pre-expression
\lean{pow_nat (add (neg one) x) (bit0 one)},
which elaborates to the expression
%\vspace{-.3em}
\begin{lstlisting}[language=lean]
pow_nat real real_has_pow_nat (add real real_has_add (neg real real_has_neg (one real real_has_one) x) (bit0 nat nat_has_add one nat nat_has_one) : real.
\end{lstlisting}
%\vspace{-.3em}

Formatted with standard notation and implicit arguments hidden, we have constructed the term
\lean{x : real ⊢ (x + -1)^2 : real}
as desired.

\subsection{Translating binding expressions}
Lean's expression structure uses anonymous bound variables to implement its \lean{pi}, \lean{lam}, and \lean{elet} binder constructs. Mathematica, in contrast, has no privileged notion of a binder. The Lean pre-expression 
$\lambda$ \lean{x, x + x}
is analogous to the Mathematica expression \mm{Function[x, x+x]}, but the underlying representation of the latter is an application of the \mm{Function} head symbol to two arguments, the symbol \mm{x} and the application expression \mm{Plus[x, x]}. Structurally it is no different from \mm{List[x, x+x]}.

To properly interpret binder expressions, both translation routines need a notion of an environment. We extend the Mathematica function \mm{LeanForm} with another argument, a list of symbols \mm{env} tracking binder depth. When the translation routine encounters a binding expression, it creates a new symbol, prepends it to the \mm{env}, and translates the binder body under this extended environment; a bound variable \mm{LeanVar[i]} is interpreted as the $i$th entry in \mm{env}.

In the opposite translation direction, a translation environment is a map from strings (names of symbols) to expressions, that is, \lean{trans_env := rb_map string expr}. When translating a Mathematica expression such as \mm{Function[x, x+x]}, the procedure extends the environment by mapping \mm{x} to a placeholder variable, translates the body under this extended environment, and then abstracts over the placeholder. Unlike in Lean, where \lean{pi}, \lean{lam}, and \lean{elet} expressions are the only expressions that encode binders, there are many Mathematica head symbols (e.g. \mm{Function}, \mm{Integrate}, \mm{Sum}) that must be translated this way.

\section{Connection Interface}
\label{section:interface}

Because of the cost of launching a new Mathematica kernel, it is undesirable to do so every time Mathematica is queried from Lean. Instead, we implement a simple server in Mathematica, which recieves requests containing expressions and returns the results of evaluating these expressions. Lean communicates with this server by calling a simple Python client script. This short script is the only part of the link that is implemented neither in either Lean nor in Mathematica.

This architecture ensures that a single Mathematica kernel will be used for as long as possible, across multiple tactic executions and possibly even multiple Lean projects. To preserve an illusion of ``statelessness,'' each Mathematica evaluation occurs in a new context which is immediately cleared. While this avoids accidental leaks of information, it is not a watertight seal, and users who consciously wish to preserve information between sessions can do so.

The translation procedure is exposed in Lean using the tactic framework via the declaration
%\vspace{-.3em}
\begin{lstlisting}[language=lean]
meta def mathematica.execute : string → tactic mmexpr.
\end{lstlisting}
%\vspace{-.3em}
This tactic evaluates the input string in Mathematica, and returns a term with type \lean{mmexpr} representing the result of the computation. From this basic tactic, it is easy to define variants such as
%\vspace{-.3em}
\begin{lstlisting}[language=lean]
run_command_using : (string → string) → expr → string → tactic pexpr.
\end{lstlisting}
%\vspace{-.3em}
The first argument is a Mathematica command, including a placeholder bound variable, which is replaced by the Mathematica representation of the \lean{expr} argument. The \lean{string} argument is the path to a file which contains auxiliary definitions, usable in the command. This variant will  apply the back-translation \lean{pexpr_of_mmexpr} to produce a \lean{pexpr}.

Another variant, \lean{execute_global : string → tactic mmexpr}, evaluates its input in Mathematica's global context.

Going back to our running example from Section \ref{section:translation}, assuming \lean{e} is the unfactored expression, we would call
%\vspace{-.3em}
\begin{lstlisting}[language=lean]
run_command_on (λ s, s ++ " // LeanConvert // Activate // Factor") e
\end{lstlisting}
%\vspace{-.3em}
to produce a pre-expression representing the factored form of \lean{e}. (Recall that the Mathematica syntax \mm{x // f} reduces to \mm{f[x]}.)
In fact, we can define 
%\vspace{-.3em}
\begin{lstlisting}[language=lean]
meta def factor (e : expr) : tactic pexpr := 
run_command_on (λ s, s ++ " // LeanConvert // Activate // Factor") e, 
\end{lstlisting}
%\vspace{-.3em}
or a variant that elaborates the result into an \lean{expr} with the same type as \lean{e}.

\section{Verification of results}
\label{section:verification}

So far we have described how to embed a Lean expression in Mathematica, manipulate it, and import the result back into Lean. At this point, the imported result is simply a new expression: no connection has been established between the original and the result. In our factoring example, we expect the two expressions to be equal; if we were computing an antiderivative, we would expect the derivative of the result to be equal to the original. More complex return types can lead to more complex relations. For example, an algorithm using Mathematica's linear arithmetic tools to verify the unsatisfiability of a system of equations may return a certificate that must be converted into a proof of falsity.

Credulous users may simply decide to trust the translation and CAS computation, and assert without proof that the result has an expected property. 
An example using this approach is given at the end of this section.
Of course, the level of trust needed to do this is unacceptably high for many situations. We are often interested in performing \emph{certifiable} calculations in Mathematica, and using this certificate to construct proofs in Lean. 

It would be hopeless to expect one tool to verify all results. Rather, for each common computation, we will have a tactic script to (attempt to) prove the appropriate relation between input and output. ``Uncommon'' or one-off computations can be verified in-line by the user. This method of separating search (or computation) and verification is discussed at length by Harrison and Th\'ery \cite{Harrison1998}, and by many others. It turns out that a surprising number of algorithms are able to generate certificates to this end.

The tactics used in this section, along with more examples, are available in the supplementary information to this paper. These examples are not meant to be exhaustive, but rather to illustrate the ease with which Mathematica can be accessed; with the possible exception of the linear arithmetic tactic, each is fairly simple to implement. The Lean library is still under development, and some types and functions used here are in fact axiomatized constants, but the implementation is not relevant to the behavior of our link.

\subsection{Factoring}
\label{subsection:verification:factoring}
In our running example, we have used Mathematica to construct the Lean expression \lean{(x + -1)^2 : real}. We expect to find a proof that \lean{x^2 - 2*x + 1 = (x + -1)^2}. 
This type of proof is easy to automate with Lean's simplifier:

\begin{lstlisting}[language=lean]
meta def eq_by_simp (e1 e2 : expr) : tactic expr := 
do gl ← mk_app `eq [e1, e2],
   mk_inhabitant_using gl simp <|> fail "unable to simplify"
\end{lstlisting}

Using this machinery, we can easily write a tactic \lean{factor} that, given a polynomial expression, factors it and adds a constant to the local context asserting equality. (The theorem \lean{sq_nonneg} proves that the square of a real number is nonnegative.)

\begin{lstlisting}[language=lean]
example (x : ℝ) : x^2-2*x+1 ≥ 0 :=
by factor x^2-2*x+1 using q; rewrite q; apply sq_nonneg
\end{lstlisting}
 
We provide more examples of this tactic in action in the supplementary material, including one in which \lean{x^10-y^10} factors into 
%\vspace{-.3em}
\begin{lstlisting}[language=lean]
(x + -1 * y) * (x + y) * (x^4 + -1 * x^3 * y + x^2 * y^2 + -1 * x * y^3 + y^4) * (x^4 + x^3 * y + x^2 * y^2 + x * y^3 + y^4). 
\end{lstlisting}
%\vspace{-.3em}

In general, factoring problems are easily handled by this type of approach, since the results serve as their own certificates. Factoring integers is a simple example of this (to verify, simply multiply out the prime factors); dually, primality certificates can be checked as in Pratt \cite{Pratt1975}. 

Factoring matrices is slightly more complex. Mathematica implements a number of common matrix decomposition methods, whose computation can be verified in Lean by re-multiplying the factors. We can use these tools to, e.g., define a tactic \lean{lu_decomp} which computes and verifies the LU decomposition of a matrix.
\begin{lstlisting}[language=lean]
example : ∃ l u, is_lower_triangular l ∧ is_upper_triangular u
        ∧ l ** u = [[1, 2, 3], [1, 4, 9], [1, 8, 27]] := by lu_decomp
\end{lstlisting}

\vspace{-1em}
\subsection{Solving polynomials}

Mathematica implements numerous decision procedures and heuristics for solving systems of equations. Many of these are bundled into its \mm{Solve} function. Over some domains, it is possible to verify solutions in Lean using the simplifier, arithmetic normalizer, or other automation. Lean's \lean{norm_num} tactic, which reduces arithmetic comparisons between numerals, is well-suited to verifying solutions to systems of polynomial equations. The tactic \lean{solve_polys} uses \mm{Solve} and \lean{norm_num} to prove theorems such as 
\begin{lstlisting}[language=lean]
example :  ∃ x y : ℝ, 99/20*y^2 - x^2*y + x*y = 0 
  ∧ 2*y^3 - 2*x^2*y^2 - 2*x^3 + 6381/4 = 0 := by solve_polys.
\end{lstlisting}

Users familiar with Mathematica may recall that \mm{Solve} outputs a list of lists of applications of the \mm{Rule} symbol, each mapping a variable to a value. A \mm{Rule} has no close correspondent in Lean, and it would involve some contortion to translate this output and extract a single solution in the proof assistant. However, it is easy to perform this transformation within Mathematica, and processing the result of \mm{Solve} \emph{before} transporting it back to Lean makes the procedure much simpler to implement. This type of consideration appears often: some transformations are more easily achieved in one system or the other. 

%\subsection{Antiderivation}
%Derivatives can be automatically computed over certain classes of real-valued functions; it is often much harder to compute antiderivatives (or indefinite integrals). We will focus on polynomials here for brevity, although the procedure is possible (and more useful) over e.g. trigonometric functions.
%
%Mathematica has a highly developed library of integration tools. We can verify that it has correctly computed an antiderivative by deriving the result and showing that it is equal (via function extensionality) to the original function. The Lean 3 library does not yet contain real analysis, but we can axiomatize the theory and use the following tactic:
%
%[FILL IN DERIVATIVE TACTIC]
\vspace{-.5em}
\subsection{Linear arithmetic}
Many proof assistants provide tools for automatically proving linear arithmetic goals, or equivalently for proving the unsatisfiability of a set of linear hypotheses. There are various techniques for doing this, including building proof terms incrementally using Fourier--Motzkin elimination \cite{Williams1986}. Alternatively, linear programming can be used to generate certificates of unsatisfiability. In this setting, a certificate for the unsatisfiability of $\{p_i(\bar x) \leq 0 : 0 \leq i \leq n \}$ is a list of rational coefficients $\{c_i : 0 \leq i \leq n\}$ such that
$\sum_{0\leq i \leq n} c_i \cdot p_i = q > 0$
for some constant polynomial $q$; equivalently, this list serves as a witness for Farkas' lemma \cite{schrijver:86}.

Given a set of hypotheses in Lean that express linear inequalities, we can prove their unsatisfiability by generating a list of such coefficients (in Mathematica), automatically proving (in Lean) that these coefficients have the necessary properties, and applying a verified proof of Farkas' lemma. 

While passing a list of inequalities to Mathematica may seem different than passing an expression such as \lean{x^2 - 2*x + 1}, we are able to use the same translation procedure. The expression \lean{x + 1 ≤ 2*y} has type \lean{Prop}, which is to say it is a type living in the lowest universe level \lean{Sort 0}. A term of this type is a proof of the claim \lean{x + 1 ≤ 2*y}. In our factorization example, we translated a term of type \lean{real}, whereas here we translate the \emph{type} of a hypothesis. But in dependent type theory, types are terms themselves, and we are able to represent any term in Mathematica. In Lean we define
%\vspace{-.3em}
\begin{lstlisting}[language=lean]
le {u} : Π {A : Type u} [has_le A], A → A → Prop.
\end{lstlisting}
%\vspace{-.3em}
We reduce this in Mathematica using the rule
%\vspace{-.3em}
\begin{verbatim}
LeanForm[LeanApp[LeanApp[LeanApp[LeanApp[LeanConst["le",
  _], _], _], x_], y_]] = Inactive[LessEqual][x, y]
\end{verbatim}
%\vspace{-.3em}
and define similar rules for $<$, $\ge$, $>$, and $=$.

Once the hypotheses have been translated to Mathematica, we must set up and solve the appropriate linear program. (Note that we are not trying to solve the hypotheses as given, but rather to find a certificate of their unsatisfiability.) A program provided in the supplementary materials to this paper shows how to use the Mathematica function \mm{FindInstance} to produce the desired list of rational coefficients. This list is translated back to Lean, where it can be elaborated with type \lean{list rat}. Once this list is confirmed to meet the requirements of Farkas' lemma, the lemma is applied to produce a proof of false. 

\begin{lstlisting}[language=lean]
example (x y : ℝ) (h1 : 2*x + 4*y ≤ 4) (h2 : -x ≤ 1) 
        (h3 : -y ≤ -5) : false := 
by not_exists_of_linear_hyps h1 h2 h3
\end{lstlisting}

\subsection{Sanity checking}

Even non-certifiable computations can sometimes be useful for proof assistant users. The \mm{FindInstance} function, for example, can be used to check that a goal is in fact provable. We define a tactic \lean{sanity_check}, which fails if Mathematica is able to find a variable assignment that satisfies the local hypotheses and the negation of the current goal. The first example below fails when Mathematica decides that the goal does not follow; the second succeeds.
%\vspace{-.3em}
\begin{lstlisting}[language=lean]
example (x : ℝ) (h1 : sin x = 0) (h2 : cos x > 0) : x = 0 :=
by sanity_check; admit

example (x : ℝ) (h1 : sin x = 0) (h2 : cos x > 0) 
             (h3 : -pi < x ∧ x < pi) : x = 0 :=
by sanity_check; admit
\end{lstlisting}

\vspace{-1em}

\subsection{Axiomatized computations}

Since it is possible to declare axioms from within the Lean tactic framework, we can axiomatize the results of Mathematica computations dynamically. This allows us to access a wealth of information within Mathematica, at least when we are not concerned about complete verification. One interesting application is to query Mathematica for special function identities. While these identities may be difficult to formally prove, trusting Mathematica allows us to find some middle ground. The \lean{mk_bessel_eq} tactic uses Mathematica's \mm{FullSimplify} function to reduce the Bessel function expression on the left, and after checking that it is equal to the one on the right, adds this equality as an axiom in Lean:
%\vspace{-.3em}
\begin{lstlisting}[language=lean]
example : ∀ x, x*BesselJ 2 x + x*BesselJ 0 x = 2*BesselJ 1 x :=
by mk_bessel_eq
\end{lstlisting}

We can also define a tactic that uses Mathematica to obtain numerical approximations of constants, and axiomatizes bounds on their accuracy:
%\vspace{-.3em}
\begin{lstlisting}[language=lean]
approx (100 * BesselJ 2 (13 / 25)) (0.001 : ℝ) 
\end{lstlisting}
%\vspace{-.3em}
declares an axiom stating that
%\vspace{-.3em}
\begin{lstlisting}[language=lean]
75977 / 23000 < 100 * BesselJ 2 (13 / 25) < 76023 / 23000.
\end{lstlisting}

\section{Concluding thoughts}
\label{section:conclusion}

\subsection{Related work}
\label{subsection:conclusion:related}
The following discussion is not meant to be comprehensive, but rather to indicate the many ways in which one can approach connecting ITP and computer algebra.

Harrison and Th\'ery \cite{Harrison1998} describe a ``skeptical'' link between HOL and Maple that follows a similar approach to our bridge. Computation is done in a standard, standalone version of the CAS, and sent to the proof assistant for certification. The running examples used are factorization of polynomials and antiderivation. The discussion is accompanied by an illuminating comparison between proof search and proof checking, and the relation to the class NP. Delahaye and Mayero \cite{Delahaye2005} provide a similar link between Coq and Maple, specialized to proving field identities.

Ballarin and Paulson \cite{Ballarin1999} provide a connection between Isabelle and the computer algebra library $\Sigma^\text{IT}$ \cite{bronstein1996} that is more trusting than the previous approach. They distinguish between sound and unsound algorithms in computer algebra: roughly, a sound algorithm is one whose correctness is provable, while an unsound algorithm may make unreasonable assumptions about the input data. 
%(Note that the term ``ad hoc'' is used in a somewhat different sense here than in the title of this paper.) 
Their link accepts sound algorithms in the CAS as oracles. A similarly trustful link between Isabelle and Maple, by Ballarin, Homann, and Calmet \cite{Ballarin1995} allows the Isabelle user to introduce equalities derived in the CAS as rewrite rules. A third example by Seddiki, Dunchev, Khan-Afshar, and Tahar \cite{Seddiki2015} connects HOL Light to Mathematica via OpenMath, introducing results from the CAS as HOL axioms.

A related, more skeptical, approach is to formally verify CAS algorithms and incorporate them into a proof assistant via reflection. This approach is taken by D\'en\`es, M\"ortberg, and Siles \cite{Denes2012}, whose CoqEAL library implements a number of algorithms in Coq.

Kerber, Kohlhase, and Sorge \cite{kerber:1998} describe how computer algebra can be used in proof assistants for the purpose of proof planning. They implement a minimal CAS, which is able to produce high-level sketch information. This sketch can be processed into a proof plan, which can be further expanded into a detailed proof.

Alternatively, one can build a CAS inside a proof assistant without reflection, such that proof terms are carried through the computation. Kaliszyk and Wiedijk \cite{Kaliszyk2007} implement such a system in HOL Light, exhibiting techniques for simplification, numeric approximation, and antiderivation.

Going in the opposite direction, CAS users may want to access ATP or ITP systems. One example of a link in this direction is Adams et al. \cite{Adams2001}, who use PVS to verify side conditions generated in computations in Maple. Systems such as Analytica \cite{Bauer1998} and Theorema \cite{Buchberger2016} provide ATP- or ITP-style behavior from within Mathematica. Axiom \cite{daly2005} and its related projects provide a type system for computer algebra, which is claimed to be ``almost'' strong enough to make use of the Curry--Howard isomorphism.

\subsection{Future work}
\label{subsection:conclusion:future}

There is much room for an improved interface under the current ITP--CAS relationship. We imagine a link integrated with Lean's supported editors, where the user effectively has access to the Mathematica REPL augmented by the current Lean environment. 

The server interface descriped in Section \ref{section:interface} only supports sequential evaluation of Mathematica commands. Both systems support parallel computation, and integrating the two could increase the utility of this link for large projects.

We have described a master--slave relationship between Lean and Mathematica respectively. The Wolfram Language is able to express ``propositions,'' and has some capacity for evaluating such propositions, but has no notion of proof. Reversing the relationship, so that a Mathematica user could use Lean to (automatically or interactively) verify propositions, is a promising direction for future work. The tools described in this paper, particularly the Mathematica-to-Lean translation, provide building blocks for such a project, and some preliminary work in this direction has begun.
This approach has some caveats. Since Mathematica is untyped, naive translations may be ambiguous; since there is not a well-specified logic underlying the CAS, one must worry about inconsistencies between its built-in assumptions and the logical rules of the proof assistant. There is no analogue to the ad hoc verification that we use in the original direction. Nonetheless, it is an interesting direction to explore.

%\acks
\paragraph{Acknowledgments.}
Many thanks to Jeremy Avigad, Jasmin Blanchette, Leonardo de Moura, Ian Ford, Johannes H\"olzl, Jos\'e Mart\'in-Garc\'ia, James Mulnix, Michael Trott, Minchao Wu, and the Lean working group at CMU.

%\nocite{*}
%\bibliographystyle{eptcs}
%\bibliography{citations}
\bibliography{lean_mm_pxtp.bbl}
\end{document}